\documentclass[a4paper,fleqn]{article}
\usepackage{enumerate}
\usepackage{graphicx}
\usepackage{harvard}
\usepackage{amsfonts}
\usepackage{amsbsy}
\usepackage[mathscr]{eucal}
\DeclareMathAlphabet\EuRscript{U}{eur}{m}{n}
\newcommand{\cur}{\EuRscript}
\newcommand{\curA}{\cur{A}}
\newcommand{\curB}{\cur{B}}
\newcommand{\scan}{\mathaccent"7017}
\newcommand{\bx}{\boldsymbol{x}}
\newcommand{\bm}{\boldsymbol{m}}
\newcommand{\bi}{\boldsymbol{i}}
\newcommand{\bB}{\boldsymbol{B}}
\newcommand{\btheta}{\boldsymbol{\theta}}
\newcommand{\bomega}{\boldsymbol{\omega}}
\newcommand{\fI}{\frak{I}}
\newcommand{\fR}{\frak{R}}
\newcommand{\eS}{\mathscr{S}}
\newcommand{\eF}{\mathscr{F}}
\newcommand{\eH}{\mathscr{H}}
\newcommand{\mcV}{{\mathcal{V}}}
\newcommand{\mcB}{{\mathcal{B}}}
\newcommand{\tB}{{{\mbox{\tiny B}}}}
\newcommand{\tC}{{{\mbox{\tiny C}}}}
\newcommand{\tG}{{{\mbox{\tiny G}}}}
\newcommand{\tF}{{{\mbox{\tiny F}}}}

\newcommand{\tmN}{{{\mbox{\tiny$N$}}}}
\newcommand{\dd}{{\mathrm{d}}}
\def\pairsep{\hspace{1cm}}

\begin{document}
\bibliographystyle{dcu}
\citationstyle{dcu}
\citationmode{abbr}
\title{The Spin-Echo System Reconsidered\thanks{To appear in {\em Foundations of Physics}, \textbf{34}, 669--688 (2004).}}
\author{D.\ A.\ Lavis\\
Department of Mathematics\\
King's College, Strand, London WC2R 2LS, U.\ K.\ \\
\small{Email:David.Lavis@kcl.ac.uk}}

\date{}

\maketitle

\begin{abstract}
\noindent Simple models have played an important role
in the discussion of foundational issues in
statistical mechanics. Among them the spin--echo
system is of particular interest since it can
be realized experimentally. This has led to
inferences being drawn about approaches to the
foundations of statistical mechanics, particularly
with respect to the use of coarse-graining.
We examine these claims with the help of computer
simulations.
\end{abstract}
\section{Introduction}\label{intro}
Kinetic equations are very useful in statistical mechanics
but they are, in general, approximations to the behaviour of the underlying
systems. Therefore, any conclusions which can be drawn from them are of limited
significance for the resolution of foundational issues. What are needed are `exact' results,
or at least situations in which numerical errors do not affect qualitative
behaviour. This is a severe restriction;
most interesting problems in statistical
mechanics concern cooperative systems and, even at equilibrium \citeaffixed{bax}{see e.g.}
there are few of these which can be solved exactly. So, of necessity, useful examples are of
assemblies of non-interacting microsystems and the
literature contains discussions
of many `toy models' of this kind, some stochastic and some
deterministic. Simulations for a number of these are
available in \citeasnoun{lavis3a};
here we confine our attention to an assembly of magnetic dipoles
precessing in a field. We shall investigate the time-evolution of the {\em Boltzmann entropy},
the fine-grained and coarse-grained versions of the
{\em Gibbs entropy} and the magnetization. We reverse the dynamic evolution
at an instant of time and demonstrate that the system returns to a state
equivalent to that at the initial time. This is the {\em spin--echo effect}.
\subsection{Forms of Entropy}\label{foe}
Consider a system, which at time $t$ has
a microstate given by  the vector $\bx(t)$ in the {\em phase-space}
$\Gamma$. Some autonomous dynamics $\bx\rightarrow \phi_t\,\bx$, ($t\ge 0$)
determines a {\em flow} in $\Gamma$ and the set of points
$\bx(t)=\phi_t \bx(0)$, parameterized by $t\ge 0$,  gives a
{\em trajectory}. The set of mappings $\{\phi_t\}_{t\ge 0}$
is a semi-group.
The system is {\em reversible} if there exists an idempotent operator $\fI$ on the
points of $\Gamma$, such that
$\phi_t\bx=\bx'$ implies that
$\phi_t \fI\,\bx'=\fI\,\bx$. Then
$\phi_{-t}=(\phi_t)^{-1}=\fI\phi_t\fI$ and the set $\{\phi_t\}$ with $t\in \Bbb{R}$ or $\Bbb{Z}$ is a
group.
\subsubsection{The Boltzmann Entropy}\label{tbe}
Macrostates (observable states) are defined by a set $\Xi$ of
macroscopic variables.\footnote{These may include some
thermodynamic variables (volume, number of particles etc.) but
they will also include other variables, specifying, for example,
the number of particles in a set of subvolumes. \citeasnoun{rid}
denotes these by the collective name of {\em supra-thermodynamic
variables}.} Let the set of macrostates be $\{\mu\}_\Xi$. They are
so defined that every $\bx\in\Gamma$ is in exactly one
macrostate denoted by $\mu(\bx)$ and the mapping $\bx\rightarrow
\mu(\bx)$ is many-one. Every macrostate $\mu$ is associated with
its `volume' $\mcV_\Xi(\mu)$ in $\Gamma$.\footnote{The term `volume'
being taken to mean some appropriate measure on $\Gamma$.}
 We thus have the map $\bx\rightarrow \mu(\bx)\rightarrow
\mcV_\Xi(\bx)\equiv\mcV_\Xi(\mu(\bx))$ from $\Gamma$ to $\Bbb{R}^+$ or $\Bbb{N}$. The
{\em Boltzmann entropy} is defined by
\begin{equation}
S_{\tB}(\bx)=k_{\tB}\ln[\mcV_\Xi(\bx)].\label{typ2}
\end{equation}
This is a phase function depending on the choice of macroscopic
variables $\Xi$.

Suppose the system consists of $N$ identical
microsystems.\footnote{In indication of which we
denote the phase-space by $\Gamma_{\tmN}$.}
Then $\Gamma_{\tmN}$ is the direct product of $N$
copies of $\Gamma_1$, the phase-space of one microsystem.
Let $\scan{\bx}^{(i)}(t)$ be the phase vector of the
$i$-th microsystem moving in its $\Gamma_1$. Now divide
$\Gamma_1$ into a enumerable set of cells $\gamma_k$
of equal volume $\nu$ such that every point in $\Gamma_1$
belongs to exactly one $\gamma_k$. The macroscopic variables
$\Xi$ are taken to be the set $\{N_k\}$ of {\em coarse-graining
variables}, where $N_k$
is the number of microsystems with phase-points in $\gamma_k$.
Then a macrostate is the part of $\Gamma_{\tmN}$
corresponding to a fixed set of values of $\{N_k\}$ and
\begin{eqnarray}
\mcV_{\{N_k\}}(\bx)&=& \Omega(\{N_k(\bx)\})\nu^{\tmN},\pairsep
\Omega(\{N_k\})=\frac{N!}{\prod_k(N_k)!},\label{ex1}\\
S_{\tB}(\bx)&=&k_{\tB}\ln[\Omega(\{N_k(\bx)\})]+k_{\tB}N\ln(\nu).\label{ex2}
\end{eqnarray}
This formula is valid irrespective of whether the microsystems
are interacting. However, if they are, then constraints
will apply to the possible values of $\{N_k\}$.\footnote{Representing, for
example, the condition that the phase point of the whole system
must lie on an energy hypersurface in $\Gamma_{\tmN}$.}
\subsubsection{The Gibbs Entropy}\label{tge}
The fine-grained Gibbs entropy\footnote{The `fine-grained' qualification
to the Gibbs entropy and probability density function is a convenient
distinction from the coarse-grained versions defined below.}
 is given by the functional
\begin{equation}
S_{\tF\tG\tG}[\rho_{\tmN}(t)]=-k_{\tB}\int_{\Gamma_{\tmN}}
 \rho_{\tmN}(\bx;t)\ln\{ \rho_{\tmN}(\bx;t)\}\dd \Gamma_{\tmN}.
\label{se9}
\end{equation}
of the fine-grained probability density function $ \rho_{\tmN}(\bx;t)$
on $\Gamma_{\tmN}$. For a measure-preserving
system for which $\rho_{\tmN}(\bx;t)$ satisfies Liouville's
equation $S_{\tF\tG\tG}[\rho_{\tmN}(t)]$ remains constant with time,
as we shall demonstrate explicitly for the spin system in
Sec.\ \ref{spec}. The resolution to this problem suggested by \citeasnoun[p.\ 148]{gib}
\citeaffixed{ehr2}{see also} is to coarse-grain the phase-space $\Gamma_{\tmN}$,
in the manner in which macrostates have been obtained in the Boltzmann approach.
We first note that for a system of identical non-interacting microsystems the probability
density function factorizes into a product of single-microsystem densities.
\begin{equation}
\rho_{\tmN}(\bx;t)=\prod_{i=1}^{\tmN} \rho_1(\scan{\bx}^{(i)};t).\label{se9a}
\end{equation}
Then
\begin{equation}
S_{\tF\tG\tG}[\rho_{\tmN}(t)]=-k_{\tB}N\int_{\Gamma_1}
 \rho_1(\scan{\bx};t)\ln\{ \rho_1(\scan{\bx};t)\}\dd \Gamma_1.
\label{se9b}
\end{equation}
Using the cells $\gamma_k$ defined in Sec. \ref{tbe} we  define the coarse-grained
probability density by
\begin{equation}
\tilde{\rho}_1(k;t)=
\int_{\gamma_k}\rho_1(\scan{\bx};t) \dd\Gamma_1
\label{se17}
\end{equation}
and the coarse-grained Gibbs entropy by
\begin{equation}
S_{\tC\tG\tG}[\tilde{\rho}_{\tmN}(t)]=-k_{\tB}N
\sum_k\tilde{ \rho}_1(k;t)
\ln\{ \tilde{\rho}_1(k;t)\}+k_{\tB}N\ln(\nu).
\label{se18}
\end{equation}
The second term in (\ref{se18}) is required for
consistency with the fine-grained entropy in the case where
the fine-grained density is uniform (with possibly different values)
over each of the cells. Then, from (\ref{se17}),
$\tilde{\rho}_1(k;t)=\nu\rho_1(\scan{\bx}_k;t)$, where $\scan{\bx}_k$
is any point in $\gamma_k$ and substituting into (\ref{se9b})
gives (\ref{se18}).\footnote{Alternatively the final term in (\ref{se18}) could be
absorbed if the formula were written in the form of an integral (rather than summation)
over the piecewise constant coarse-grained density.}

If we begin with any
fine-grained density $\rho_{\tmN}(\bx;t)$ and calculate
$S_{\tF\tG\tG}[\rho_{\tmN}(t)]$, and then apply coarse-graining
and calculate $S_{\tC\tG\tG}[\tilde{\rho}_{\tmN}(t)]$,
\begin{equation}
S_{\tF\tG\tG}[\rho_{\tmN}(t)]\le
S_{\tC\tG\tG}[\tilde{\rho}_{\tmN}(t)],\label{se18b}
\end{equation}
with equality only if the fine-grained density is uniform over the cells
of the coarse-graining. Now we can conceive of two possible ways of tracing
the evolution of entropy in the Gibbs coarse-grained picture.
\begin{enumerate}[(i)]
\item We could begin with some fine-grained density giving entropy
$S_{\tF\tG\tG}[\rho_{\tmN}(0)]$
at $t=0$ and watch its evolution as time increases. If at time $t'\ge 0$ we coarse-grain, then
\begin{equation}
S_{\tF\tG\tG}[\rho_{\tmN}(0)]
=S_{\tF\tG\tG}[\rho_{\tmN}(t')]
\le S_{\tC\tG\tG}[\tilde{\rho}_{\tmN}(t')].
\label{18c}
\end{equation}
However if we coarse-grain at two instants $0\le t'< t''$ it is not
necessarily the case that
\begin{equation}
S_{\tC\tG\tG}[\tilde{\rho}_{\tmN}(t')]
\le S_{\tC\tG\tG}[\tilde{\rho}_{\tmN}(t'')].
\label{18d}
\end{equation}
The coarse-grained entropy will not necessarily show monotonic increase.
However, the graph of the coarse-grained entropy will not depend on the
instants at which coarse-graining is applied.
\item If, instead of the strategy adopted in (i) we coarse-grain at $t'$
then follow the evolution of the coarse-grained density and
then {\em re-coarse-grain} at the later time $t''$,  (\ref{18d})
will hold. Course-grained entropy will show monotonic increase. However,
the graph of entropy against time will be affected by the instances at which
coarse-graining is applied.
\end{enumerate}

\vspace{0.5cm}

\noindent From (\ref{ex1})--(\ref{ex2}), using Stirling's formula for large $N$,\footnote{In fact
the approximation is close only when not only $N$, but all the $N_k$ are large. This
means that it is good only for large $N$ and a distribution of microsystems close
to the uniform distribution over the cells.}
\begin{equation}
S_{\tB}(\bx) \simeq -k_{\tB}N\sum_k \frac{N_k(\bx)}{N}\ln\left(\frac{N_k(\bx)}{N}\right)
+k_{\tB}N\ln(\nu).\label{ex3}
\end{equation}
The relationship between (\ref{se18}) and (\ref{ex3}) is now easy to see. If on the
one hand a very large assembly of microsystems is taken with initial density in
$\Gamma_1$ of $N\rho_1(\bx;0)$ then $N_k(t)/N$, the proportion of the
assembly in cell $\gamma_k$ at time $t$ is $\tilde{\rho}_1(k;t)$ given by (\ref{se17})
and (\ref{ex3}) is asymptotically equivalent to (\ref{se18}). Conversely, if in the
Gibbs formulation the initial density function is chosen to be a set of $N$
suitably-weighted Dirac delta functions, we recover (\ref{ex3}). In summary, we expect the
Boltzmann entropy in the limit of large $N$ and close to the uniform distribution
to converge to the coarse-grained Gibbs entropy.
\section{The Model}\label{spec}
Consider the simple model in which a magnetic dipole of moment $\bm$
is fixed at its centre but is free to rotate in the presence of a constant
magnetic field $\bB$. The equation of motion of the dipole will be
\begin{equation}
\dot{\bm}(t)=g\, \bm(t) \wedge \bB,
\label{spec1}
\end{equation}
where $g$ is the gyromagnetic ratio. Released from rest the dipole will precess
at a constant angle to $\bB$.
In particular, if $\bm$ is located at the origin of a cartesian coordinate system
with $\bB$ in the direction of the negative $z$-axis and if
initially $\bm$ lies in the $x-y$ plane, its subsequent motion remains
in the $x-y$ plane and is given by
\begin{equation}
\bm(t)= (m\cos(\theta(t)),m\sin(\theta(t))),
\label{se1}
\end{equation}
where
\begin{equation}
\theta(t)=\phi_t\,\theta(0)= \eF_{2\pi}(\theta(0)+\omega t), \pairsep \omega= B\,g,
\label{se2}
\end{equation}
and\footnote{Where, of course,
 $\eF_{\alpha}(x\pm\eF_{\alpha}(y))=\eF_{\alpha}(x\pm y)$,
for all real $x$ and $y$ and positive $\alpha$.}
\begin{equation}
\eF_\alpha(x)= \alpha\times \mbox{\sf Non-Integer Part}
\left(\frac{x}{\alpha}\right).
\label{se14}
\end{equation}
Suppose that at some time $t=\tau$ the magnetic field $\bB$ is turned off and a
field $\bB'$, in the direction of the $x$--axis is turned on for a time
$t'=\pi/B'g$. The effect of this will be to rotate the dipole through an angle
$\pi$ about the $x$-axis, translating its position from
$\theta(\tau)=\eF_{2\pi}(\theta(0)+\omega\tau)$ to
$\theta'(\tau)=2\pi-\eF_{2\pi}(\theta(0)+\omega\tau)=\eF_{2\pi}(2\pi-\theta(0)-\omega\tau)$;
a reflection in the $x$-axis. We denote this idempotent reflection operator by $\fR$; that is
$\fR(\theta,\omega)=(2\pi-\theta,\omega)$. With reflection applied at $t=\tau$
\begin{equation}
\theta(2\tau)=\eF_{2\pi}(\theta'(\tau)+\omega\tau)=2\pi-\theta(0).
\label{se3}
\end{equation}
This reflectional return or {\em echo-effect} is what gives the system its name.
The model is also reversible with $\fI(\theta,\omega)=(\theta,-\omega)$. Then
\begin{equation}
\theta(2\tau)=\phi_{-\tau}\theta(\tau)=\eF_{2\pi}(\theta(\tau)-\omega\tau)=\theta(0).
\label{se3a}
\end{equation}
So the system has two mechanisms for making it `retrace its steps'. However, this
is not so strange. It would be true for any system with periodic boundary conditions;
and a similar effect occurs when a particle is in one-dimensional
motion at constant speed $v$ confined between elastic walls at $x=0$ and $x=L$.
Then we can `unfold' right-to-left motions of the particle into the
region $[L,2L]$. The model is now equivalent to the dipole motion with $\pi$ replaced
by $L$. The echo transformation $x\to 2L-x$ at $t=\tau$ is now
{\em exactly the same as} reversing the direction of the velocity, with $x(2\tau)=x(0)$
and $\dot{x}(2\tau)=-\dot{x}(0)$. However, there is a second possible
transformation $x\to L-x$. Now $x(2\tau)=L-x(0)$
and $\dot{x}(2\tau)=\dot{x}(0)$. For an assembly of particles this
fulfills the purposes of the echo transformation just as well.\footnote{It
undoes during the time interval $[\tau,2\tau]$ the spreading which has
occurred during the interval $[0,\tau]$.}

\begin{figure}[p]
\begin{flushleft}
\vspace*{-0.5cm}
\hspace*{-1.5cm}
\includegraphics[width=130mm,angle=0]{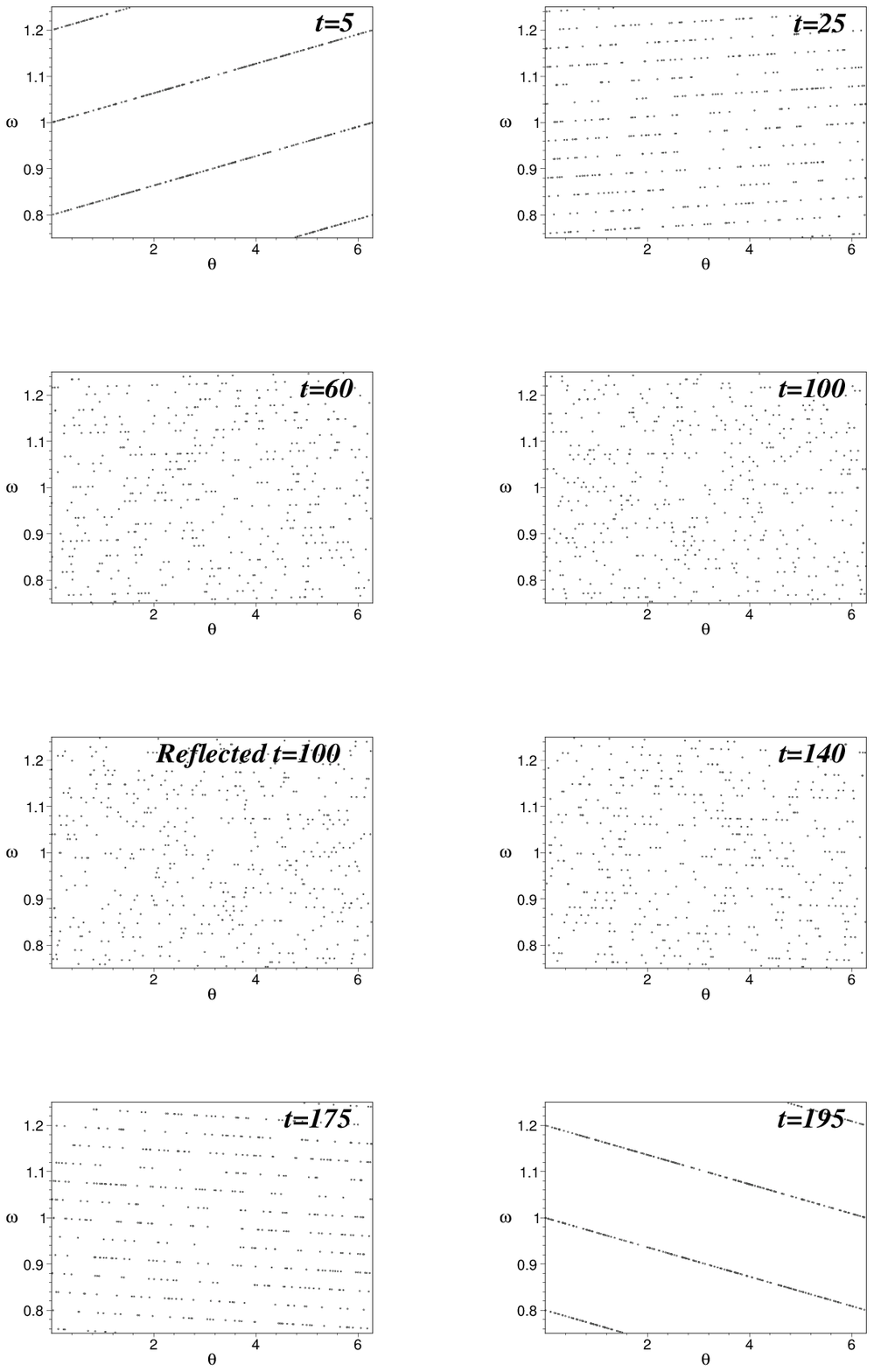}
\end{flushleft}
\caption{An assembly of $N=500$ rotating dipoles.}\label{spin-echofig1}
\end{figure}

As indicated, our interest is in an assembly of microsystems.
Consider the collection $\bm^{(i)}$, $i=1,2,\ldots,N$ of such dipoles with
angular velocities $\omega^{(i)}$ in the range $[\omega_{\rm min},\omega_{\rm max}]$
and plot their evolutions in the
$\theta-\omega$ plane. Suppose that, $N=500$, $\tau=100$,
$\omega_{\rm min}=0.75$ and $\omega_{\rm max}=1.25$ and that
 the $\omega^{(i)}$
are chosen randomly from a uniform distribution
on $[\omega_{\rm min},\omega_{\rm max}]$ with
$\theta(0)=0$ for all the dipoles. Then we have the situation shown
in Fig.\ \ref{spin-echofig1}.
At $t=0$ each dipole is aligned in the $\theta=0$ direction and at $t=5$ the
phase-points in $\Gamma_1$ form lines with this effect persisting to about $t=50$.
After this the periodic boundary conditions lead
to a breakup of the ordered appearance and a `spreading' of phase-points
in $\Gamma_1$.
When the reflectional transformation is applied at $t=\tau=100$ the distribution of phase-points
at $t>\tau$ is the mirror image in $\theta=\pi$ of its form at $2\tau-t$ and
the final configuration is along the line $\theta=2\pi$ at $t=2\tau$. A macroscopic variable
which can be used to follow the evolution of the system is the $x$ component of the magnetization
density
\begin{equation}
\sigma(t)= \frac{1}{mN}\sum_{i=1}^{\tmN} \bm^{(i)}(t)\cdot\hat{\bx}
=\frac{1}{N}\sum_{i=1}^{\tmN}\cos\left(\theta^{(i)}(t)\right).\label{ex11}
\end{equation}
\begin{figure}[t]
\begin{center}
\includegraphics[width=100mm,angle=0]{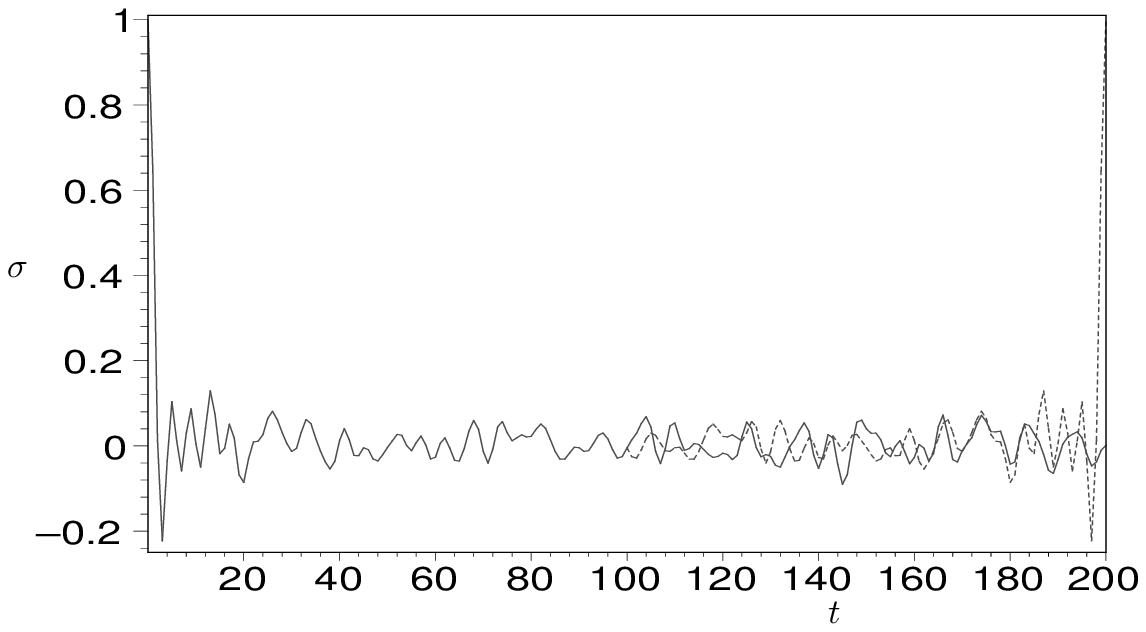}
\end{center}
\caption{The evolution of the magnetization density. After $t=\tau=100$
the broken line gives the echo.}\label{spin-echofig2}
\end{figure}
This is shown in Fig. \ref{spin-echofig2}. There is a rapid decrease of
magnetization density from its initial value of unity to fluctuations
around the perfectly spread value of $\sigma=0$. The average magnitude of these
fluctuations will be inversely proportional to $N$ and in general we expect them
to be quite small.
Since the angular
velocities have been chosen randomly the assembly is
quasi-periodic. It is also volume-preserving and will, therefore, satisfy the
\citeasnoun{poin1c} recurrence theorem. For `most' initial
points, if there in no echo reflection,  the phase point
$(\btheta,\bomega)=
(\theta^{(1)},\ldots,\theta^{(\tmN)},\omega^{(1)},\ldots,\omega^{(\tmN)})$
in the $2N$-dimensional phase-space $\Gamma_{\tmN}$ nevertheless
returns to within a neighbourhood of its initial
value.\footnote{The recurrence time will, of course, be dependent
on the size of the neighbourhood.} This will lead to a large fluctuation
in magnetization density. Of course, if the initial
angular velocities are chosen to be commensurate, the system will
be periodic and will return exactly to its initial point with $\sigma=1$.

There would be nothing particularly special about this model, if
it were not for the fact that it has been realized experimentally.
\citeasnoun{hahn1} \citeaffixed{hahn2,rhim1,brewer&hahn}{see also} applied a
magnetic field to
various liquids whose molecules contain hydrogen
atoms. By manipulating the components of
the magnetic field he was able to start with the dipole moments of
the proton spins in the $x$--direction, make them precess around the
$z$--axis and then reflect the directions of the dipoles in the
$x$--axis to achieve the echo effect with the dipoles returning to
their initial alignment.\footnote{The variations in the angular
velocities were achieved from small variations in the
strength of the magnetic field throughout the sample.} This
system has aroused some interest in relation to questions of
reversibility in statistical mechanics
\cite{blatt1,mayer&mayer,den&den,rid&red,rid}.
This will be discussed in Sec.\ \ref{conc}. Here we shall simply present
the results of our calculations.

\begin{figure}[t]
\begin{center}
\includegraphics[width=100mm,angle=0]{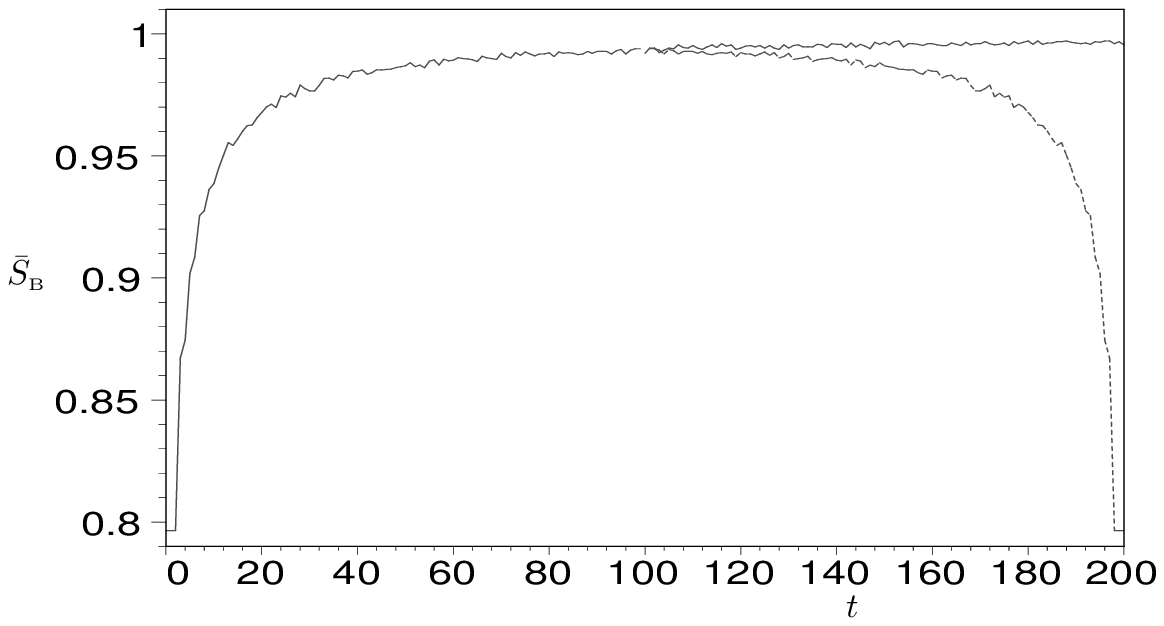}
\end{center}
\caption{The evolution of the Boltzmann entropy of the dipole assembly. After $t=\tau=100$
the broken line gives the echo. }\label{spin-echofig3}
\end{figure}

The cells to be used both for the Boltzmann entropy and the coarse-grained
Gibbs entropy are defined by
dividing the single--dipole phase-space $\Gamma_1$ into $n_\theta\times
n_\omega$ rectangles with edges parallel to the $\theta$ and $\omega$ axes
and of lengths $\triangle\theta=2\pi/n_\theta$ and
$\triangle\omega=(\omega_{\rm max}-\omega_{\rm min})/n_\omega$
respectively. In Fig.\ \ref{spin-echofig3},
we show the scaled Boltzmann entropy
\begin{equation}
\bar{S}_{\tB}(\bx(t))=\frac{S_{\tB}(\bx(t))-(S_{\tB})_{\rm min}}{(S_{\tB})_{\rm max}-(S_{\tB})_{\rm min}},
\label{ex5}
\end{equation}
for the same evolution as Fig.\ \ref{spin-echofig1}, and $n_\theta=n_\omega=100$,
where $(S_{\tB})_{\rm min}=k_{\tB}N\ln(\triangle\,\theta\triangle\omega)$ is the entropy
were all the spins to be concentrated in one cell and $(S_{\tB})_{\rm max}$
corresponds to the spins being equally distributed over the cells.\footnote{We do not,
of course, imply that these scaling factors correspond to attainable states
for the system, since the distribution of angular velocities is invariant
with time.}
The continuous and broken lines for $t>100$ correspond
respectively to the evolutions
without and with the echo-effect.
\begin{figure}[p]
\begin{flushleft}
\vspace*{-0.5cm}
\hspace*{-1cm}
\includegraphics[width=130mm,angle=0]{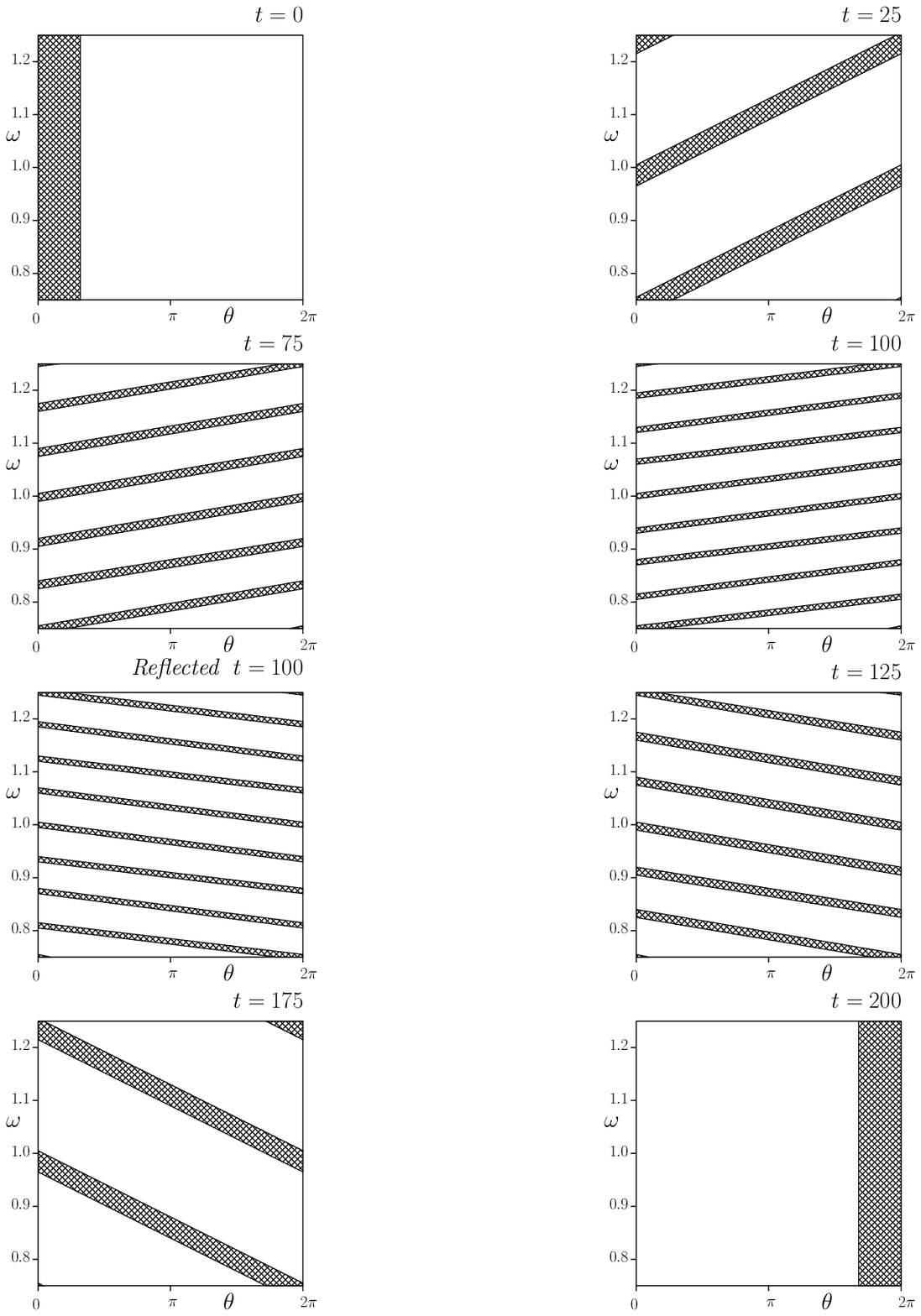}
\end{flushleft}
\caption{The evolution of the fine-grained one-dipole probability density
function with the echo occurring at $t=\tau=100$.}\label{spin-echofig4}
\end{figure}

We now calculate the fine-grained Gibbs entropy. Suppose that the initial probability
density function is concentrated and uniform over the rectangle
$\omega\in[\omega_{\rm min},\omega_{\rm max}]$, $\theta\in[0,\theta_0]$, ($\theta_0<2\pi$). Then
\begin{equation}
\rho_1(\theta,\omega;0)= \frac{\eH(\theta)-\eH(\theta-\theta_0)}{\theta_0(\omega_{\rm max}-\omega_{\rm min})},
\label{se12}
\end{equation}
where $\eH(\theta)$ is the {\em Heaviside unit function}, and
{\mathindent=0cm
\begin{equation}
\rho_1(\theta,\omega;t)=\left\{\begin{array}{l}
\displaystyle{\frac{\eH(\theta-\eF_{2\pi}(\omega t))-\eH(\theta-\eF_{2\pi}(\theta_0+\omega t))}
{\theta_0(\omega_{\rm max}-\omega_{\rm min})}},\\[0.3cm]
\hspace{6cm}\eF_{2\pi}(\omega t)<\eF_{2\pi}(\theta_0+\omega t),\\[0.4cm]
\displaystyle{\frac{\eH(\theta-\eF_{2\pi}(\omega t))-\eH(\theta-\eF_{2\pi}(\theta_0+\omega t))
+\eH(\theta)-\eH(\theta-2\pi)}
{\theta_0(\omega_{\rm max}-\omega_{\rm min})}},\\[0.3cm]
\hspace{6cm}\eF_{2\pi}(\theta_0+\omega t)<\eF_{2\pi}(\omega t).
\end{array}
\right.
\label{se15}
\end{equation}
If} the echo transformation $\theta\to 2\pi-\theta$ is applied at the time $\tau$
the one-spin probability density function for $t>\tau$ is given, in terms of
(\ref{se15}) by $\rho_1(2\pi-\theta,\omega;2\tau-t)$.
The evolution of this fine-grained
probability density function, with $\tau=100$, is shown in Fig.\ \ref{spin-echofig4}.
Over the time interval $[0,\tau]$ the cross-hatched region spreads itself in ever-thinner
striations over $\Gamma_1$ and this process would continue if the echo transformation
were not applied.\footnote{However,
we have to be a little cautious about this
since we are considering a collection of non-interacting dipoles.
For each dipole the second equation of motion to pair with
(\ref{se2}) is $\omega(t)=\omega(0)$. Motion is horizontal in
$\Gamma_1$ and, unlike for example a gas of particles moving
according to the baker's transformation \cite{lavis3a}, and
contrary to the assertion by \citeasnoun[p.\ 1248]{rid&red}
the system is {\em mixing} in $\Gamma_1$ only in a limited sense.}
The effect of the echo-transformation is as in Fig. \ref{spin-echofig1}; it
produces a configuration at $t>\tau$ which is the reflection in $\theta=\pi$ of
the configuration at $2\tau-t$.
\begin{figure}[h]
\begin{center}
\includegraphics[width=100mm,angle=0]{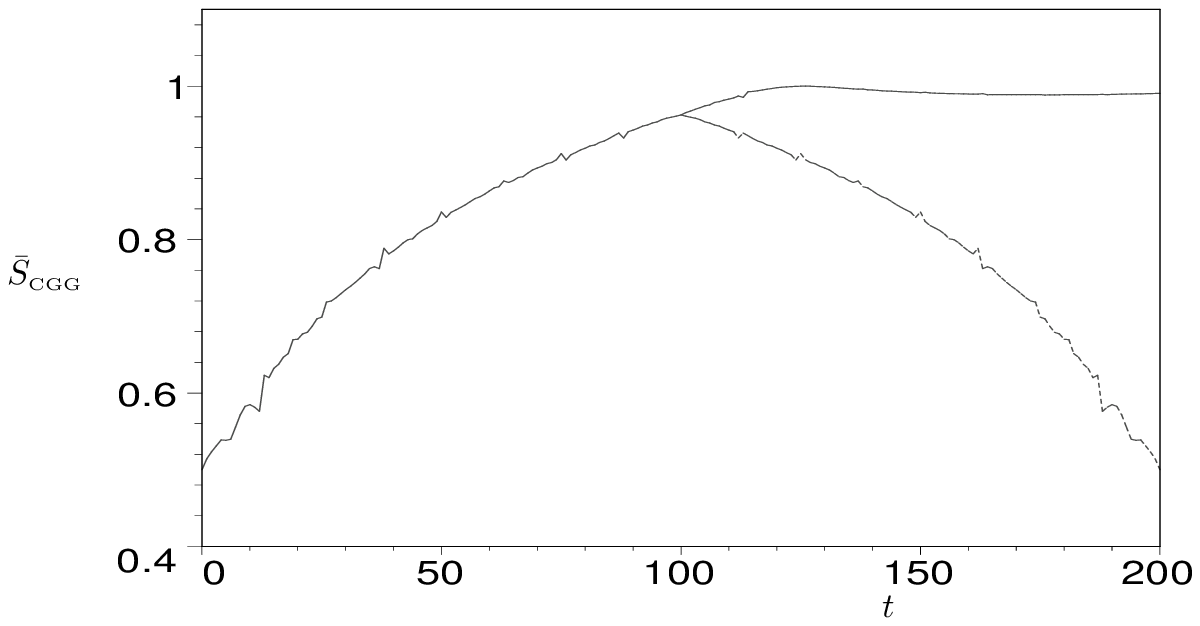}
\end{center}
\caption{The evolution of the coarse-grained Gibbs entropy of the dipole assembly. After $t=\tau=100$
the broken line corresponds to the echo. }\label{spin-echofig5}
\end{figure}
Substituting from (\ref{se15}) into (\ref{se9b}) gives
\begin{equation}
S_{\tF\tG\tG}[\rho_{\tmN}(t)]=k_{\tB}N\ln\{\theta_0(\omega_{\rm max}-\omega_{\rm min})\}.
\label{se16}
\end{equation}
This is simply an expression of the well-known result that the fine-grained Gibbs entropy is invariant
with respect to time. The coarse-grained Gibbs entropy is now calculated using the same coarse-graining as was used
to obtain the macrostates for the Boltzmann entropy.
$S_{\tC\tG\tG}[\tilde{\rho}_{\tmN}(t)]$ will have a maximum value when the cross-hatched area in Fig.\ \ref{spin-echofig4}
is spread evenly over the cells. Then
$\tilde{\rho}_1(k;t)=(\triangle\theta\triangle\omega)/\{2\pi(\omega_{\rm max}-\omega_{\rm min})\}$.
Substituting into (\ref{se18}) (with $\nu=\triangle\theta\triangle\omega$) gives
\begin{equation}
(S_{\tC\tG\tG})_{\rm max}=k_{\tB}N\ln\{2\pi(\omega_{\rm max}-\omega_{\rm min})\}.
\label{se18z}
\end{equation}
We adopt the strategy (i) of Sec.\ \ref{tge} and coarse-grain
the fine-grained density as time evolves (rather than performing successive re-coarse-grainings). The
results for $\bar{S}_{\tC\tG\tG}[\tilde{\rho}_{\tmN}(t)]=S_{\tC\tG\tG}[\tilde{\rho}_{\tmN}(t)]/(S_{\tC\tG\tG})_{\rm max}$,
when $n_\theta=n_\omega=100$, are shown in Fig.\ \ref{spin-echofig5}.

\citeasnoun{rid&red} have shown  that the coarse-grained entropy
tends to its maximum value (\ref{se18z})
as $t\to\infty$ and our simulations in Fig.\ \ref{spin-echofig5}
support this result.
\section{Discussion}\label{conc}
The first discussion of the spin--echo effect in relation to coarse-graining is
due to \citeasnoun{blatt1}. His argument is that ``the coarse-graining approach
depends crucially upon the assertion that `fine-grained' measurements are impracticable,
and thus [that] the fine-grained entropy is a meaningless concept'' (p.\ 746). Since
a counter-example to this is provided by the spin--echo system which shows that
``macroscopic observers are not restricted to coarse-grained experiments''
he concludes that it ``is not permissible to base fundamental arguments in statistical
mechanics on coarse-graining'' (p.\ 749). So what is the weight of this argument? It is based
on ingenious experiments which allow a system of independent microsystems to be
returned, by macroscopic means, to a phase state close to the one they were in
at an earlier time. Two effects could account for `closeness' rather than
exact return. The first would be be an internal cooperative effect, in this case
a spin--spin coupling.\footnote{Similar experiments including dipolar
coupling were performed by \citeasnoun{rhim1}. Whilst these are of importance
expermentally they do not affect the argument.} \citeasnoun[p.\ 750]{blatt1} remarks that the decrease
in the echo--pulse arising from this is ``from [the] present point of view accidental''.
He is content to consider a system of independent microsystems, because in any event
the inclusion of cooperative effects would not allow an escape from the iron hand
of Liouville's theorem; the fine-grained Gibbs entropy would still be constant.
He is interested in the external (spin--lattice) source of the deviation from exact return.
This interventionist alternative to coarse-graining, which is also the position of
\citeasnoun{rid&red} and \citeasnoun{rid} will not be discussed here. Rather we return
to the original contention that the demonstration of a system which can be
controlled more-or-less exactly at the microscopic level by macroscopic means is the
death-blow for coarse-graining. Of course, the coarse-graining referred to by
\citename{blatt1} (and also by \citename{rid&red} and \citename{rid}) is of
the Gibbs--Ehrenfest type and it is true that \citeasnoun[p.\ 167]{tol} in
justifying this argues that ``in making any actual measurement of the
[macroscopic variables] of the system $\ldots$ we ordinarily do not achieve
the precise knowledge of their values theoretically permitted by classical mechanics''.
But if this were the main argument for coarse-graining of the Gibbs--Erhenfest or
Boltzmann kind it would be very weak. It has always been possible to obtain
analytic solutions for assemblies of non-interacting microsystems and with
the advent of fast computing we can, as we have here, produce data for assemblies
of arbitrary size. The fact that such a system can be realized experimentally
and controlled macroscopically may have been of great importance technically,
but it is hardly a milestone in foundational development. In fact it is not clear
that either \citeasnoun{gib} or \citeasnoun{ehr2} intended to justify the procedure
by an appeal to the limitations of measurement. Gibbs (ibid, p.\ 148) refers
to the cells of the coarse-graining as being ``so small that [the fine-grained
probability density function] may in general be regarded as sensibly constant
within any one of them at the initial moment'' and the Ehrenfests (ibid, p.\ 52)
simply observe that the cells must be ``small, but finite''. In the case of
the Boltzmann entropy the situation is somewhat clearer. The size of the cells
defines the `macro-scale' as distinct from the `micro-scale' \cite{leb1}. Of course,
this demarcation is to some extent arbitrary, but it is equally so for any
macroscopic physical theory.\footnote{See e.g. the definition of fluid
density in \citeasnoun[p.\ 1]{lan&lif1}.} As is pointed out by \citeasnoun{grun1},
Boltzmann's entropy can be regarded as a measure of homogeneity and
in this context the equilibrium state corresponds simply to the maximum
entropy state, which has the most homogeneity. It is precisely and
only here, in defining a measure for homogeneity at equilibrium
\cite{rid}, that the demarcation between the macro- and micro-scales
must be made. And this is unavoidable since no distribution of discrete
points over a continuum is uniform on all scales.

\begin{figure}[t]
\begin{center}
\includegraphics[width=100mm,angle=0]{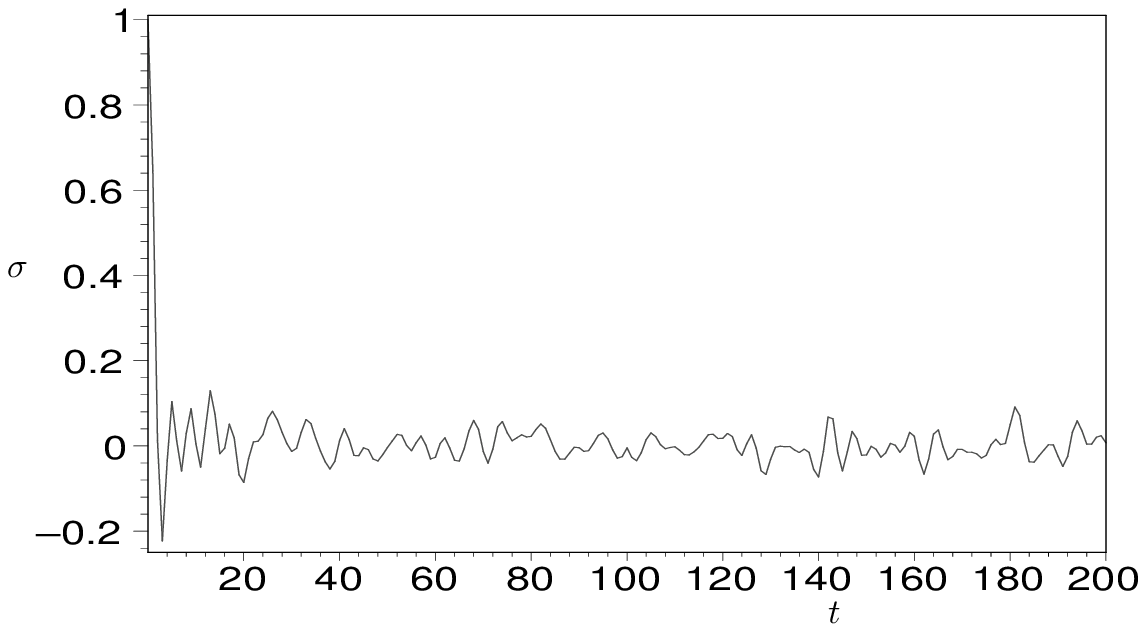}
\end{center}
\caption{The evolution of the magnetization density. After $t=100$
the system is reflected and the angular velocities are subject to small perturbations.}\label{spin-echofig6}
\end{figure}

We now consider the case made by
\citeasnoun[p.\ 49--50 and 140--143]{den&den}\footnote{The same argument is
reproduced in \citeasnoun[p.\ 1253--1254]{rid&red}.} for the
assertion that the spin--echo system exemplifies circumstances that
are ``highly exceptional'' in reproducing the kind of reversible situation used
by \citeasnoun{los1} in his challenge to Boltzmann. The argument hinges on a comparison
between a gas expanding in a box and the spin--echo system. This already presents
some problems since, as we have shown in Sec.\ \ref{spec}, the states of a particle
moving in one dimension between perfectly reflecting barriers are isomorphic to those
of a single spin precessing in a field with the spin--echo reflection
equivalent to velocity inversion. It follows from this that non-interacting
assemblies of each of these are isomorphic.\footnote{It may be that \citename{den&den}
are effectively arguing that a two-dimensional gas of particles in a box
$\mcB=\{(x,y)|0\le x\le L,0\le y\le L\}$ where each particle moves at constant speed in the $x$-direction
without any collisions is itself ``highly exceptional''. If so the spin--echo
system is irrelevant, except in so far as it is realized experimentally.}
A summary of the situation considered by \citeasnoun[p.\ 49--50]{den&den} is as
follows:\footnote{They begin by referring to a gas of particles in a box
where the operation needed to make the system retrace its steps is velocity
reversion.}
\begin{enumerate}[(i)]
\item Let $\curA\rightarrow\curB$ be a ``{\em macroscopic} process'' from a thermodynamic
state $\curA$ to a thermodynamic state $\curB$.
\item Let $\eS(\curA)$ and $\eS(\curB)$ be
``those sets of exactly specified microstates which are accessible to the gas''
in states $\curA$ and $\curB$.
\item Let $\fI{\eS}(\curA)$ and $\fI{\eS}(\curB)$
be those sets of macrostates obtained from ${\eS}(\curA)$ and ${\eS}(\curB)$
by reversing the velocities.
\item If $\bx(0)\in{\eS}(\curA)$ and $\bx(\tau)\in{\eS}(\curB)$
then $\fI\bx(\tau)\in\fI{\eS}(\curB)$ and $\phi_\tau\fI\bx(\tau)=\fI\bx(0)\in\fI{\eS}(\curA)$.
\end{enumerate}
The inference is drawn that, if the system during the evolution $\bx(0)\to\bx(\tau)$ goes from $\curA$
to $\curB$, there is an allowed evolution $\fI\bx(\tau)\to \fI\bx(0)$, taking the system
from $\curB$ to $\curA$.\footnote{There is one benign gap in this argument.
It is assumed that the thermodynamic states for the reversed process are the same as those
for the forward process. This is equivalent to supposing that, if $\bx\in{\eS}(\curA)$,
then $\fI\bx\in{\eS}(\curA)$. In other words, $\fI{\eS}(\curA)\equiv{\eS}(\curA)$,
$\fI{\eS}(\curB)\equiv{\eS}(\curB)$. The truth of these identities, although plausible, will,
of course, depend on the meaning (yet to be discussed) of  `accessible'.}
If thermodynamic entropy increases in one direction it will decrease
in the other. This is the heart of Loschmidt's paradox.
In his reply to Loschmidt, \citeasnoun{boltz3c} pointed out that, whereas the
trajectories from the majority of the points in ${\eS}(\curA)$ will yield an increase
in entropy in the time interval $[0,\tau]$, only a small percentage of the points
in $\fI{\eS}(\curB)$ will yield trajectories giving a decrease in entropy over $[0,\tau]$.
\citeasnoun[p.\ 50]{den&den} accept the general validity of this argument,
but they believe that the spin--echo system where velocity inversion $\fI$ is replaced by
reflection $\fR$ is a special case.
They claim (translating into our notation) that ``the situation
[in the spin--echo system] is that the set of the type [$\fR{\eS}(\curB)$]
contains the same number of members as the set of type [${\eS}(\curA)$]; for
every original spin there is a spin with a reversed velocity of precession''.\footnote{See also
\citeasnoun[p.\ 1254]{rid&red} for a similar assertion.} This statement contains two
parts the first contentious and the second obviously true. It is certainly true that
to every spin state there is another with the velocity of precession reversed (or the position reflected).
A similar statement would be true for {\em any} reversible dynamic system. The distinguishing,
although possibly not unique, feature of the spin system is that ``these velocities can
actually be reversed simultaneously by applying a magnetic pulse''. But this is a technical
feature which could always be anticipated for a system of non-interacting microsystems.
On the other hand if the first part of the statement (that $\fR{\eS}(\curB)$
contains the same number of members as ${\eS}(\curA)$)
were true this would be in conflict with Boltzmann's answer to Loschmidt
and it would be necessary to give an argument why this does not contradict the second
law.\footnote{Such an argument (again repeated by \citeasnoun{rid&red}) was provided by
\citeasnoun[p.\ 136]{mayer&mayer}.} The problem with understanding this argument is
in interpreting the term `accessible'.\footnote{\citeasnoun{rid&red} use the term `available'
rather than `accessible'.} Let us suppose that it is to be interpreted as {\em all those
microstates compatible with a given value for the $x$ component of the magnetization
$mN\sigma(t)$}.\footnote{The argument could be suitably modified for variants on this
definition, including a Boltzmann-like account base on macrostates.}  If initially
$\sigma(0)=1$, then all the spins must be aligned with the $x$-axis;
$\btheta(0)=\left(\theta^{(1)}(0),\ldots,\theta^{(\tmN)}(0)\right)=\bf{0}$
\, and the microstates ${\eS}(\curA)$ accessibly to this macrostate $\curA$ correspond to
all possible values of $\bomega=(\omega^{(1)},\ldots,\omega^{(\tmN)})$.
Now we have to define the final state $\curB$ at time $t=\tau$. We could
simply take this to be given by $\sigma(\tau)=0$. This would, of course, imply that
$\bomega$ is constrained by the condition
\begin{equation}
\sum_{i=1}^{\tmN}\cos\left(\omega^{(i)}\tau\right)=0.
\label{ex22}
\end{equation}
This condition will eliminate most of the points in ${\eS}(\curA)$. We have seen in Fig.\ \ref{spin-echofig2}
that a typical evolution of $\sigma(t)$ starting from alignment in the $x$ direction involves
a rapid decrease followed by oscillations about $\sigma=0$. A more realistic definition of $\curB$
is that $\sigma$ lies in some small range $[-\epsilon,\epsilon]$. This replaces the condition
(\ref{ex22}) by
\begin{equation}
\left|\sum_{i=1}^{\tmN}\cos\left(\omega^{(i)}\tau\right)\right|\le \epsilon.
\label{ex33}
\end{equation}
For sufficiently large $\tau$ this condition will include `most' of the points in ${\eS}(\curA)$.\footnote{Those
excluded will mostly be points where the angular velocities are commensurate and the motion is periodic.}
Now suppose we start at a phase point in ${\eS}(\curA)$ evolving into
\begin{equation}
\theta^{(i)}(\tau)= \eF_{2\pi}\left(\omega^{(i)}\tau\right), \pairsep i=1,2,\ldots,N
\label{ex44}
\end{equation}
with
\begin{equation}
\left|\sum_{i=1}^{\tmN}\cos\left(\theta^{(i)}(\tau)\right)\right|\le \epsilon.
\label{ex55}
\end{equation}
If we now apply the reflection $\theta^{(i)}(\tau)\to 2\pi-\theta^{(i)}(\tau)$ the value of the sum
on the left of (\ref{ex44}) is unchanged. The new reflected phase point is also in
$\fR{\eS}(\curB)\equiv{\eS}(\curB)$ and, under the evolution\footnote{With $\bi=(1,1,\ldots,1)$.}
 $\phi_t(2\pi\bi-\btheta,\bomega)$,
over the further time interval $[0,\tau]$ it returns to $\fR{\eS}(\curA)\equiv{\eS}(\curA)$.
Most of the points of ${\eS}(\curA)$ satisfy this account, but the crucial question is whether
in passing through ${\eS}(\curB)$ they include all (or even most) of the points of that set.
The answer is clearly `no'. To see this simply take a reflected point $(2\pi\bi-\btheta,\bomega)$
which does return to ${\eS}(\curA)$ and apply any one of an infinity of small perturbations to
the angular velocity. Most of these will not return to ${\eS}(\curA)$ in a time $\tau$, or in fact
in any time interval much less than the Poincar\'e recurrence time.\footnote{And even then we should
need to broaden our definition of $\curA$ around $\sigma=1$.} This situation is shown in Fig.\ \ref{spin-echofig6}.

In his account of the spin--echo system \citeasnoun[p.\ 221]{skl} comments that
\begin{quote}
 ``It is as if we could
prepare a gas in such a way that an ensemble of gases so prepared would initially be uniformly
spread throughout a box. But the overwhelming majority of the gases in the ensemble would then
spontaneously flow to the left-hand half of the box.''
\end{quote}
The problematic word in this quote is `prepare'. To prepare `from scratch' a spin
system or any other assembly is such a way that it will achieve a particular macrostate (low entropy,
high magnetization, etc.) after a particular interval of time would involve careful adjustment of the
relationships of velocities and positions for each microsystem; a task worthy of a {\em Maxwell demon}. However,
what we have here is a much simpler process. We allow the system to achieve values which imply a
recent memory of the required macrostate and then apply a reflection. This macroscopic operation by
a {\em Loschmidt demon}\footnote{The term {\em Loschmidt demon}, seems to have
been introduced by \citeasnoun{rhim1}.}
is only part of the process of preparation. The difficult part is left to the system.

The aim of the work of \citeasnoun{rid&red} is to use an examination of the spin--echo
system to discredit the use of the Gibbs-Ehrenfest coarse-graining in favour of an
interventionist approach. While it is true that the status of the coarse-grained
Gibbs entropy lacks the clarity of the Boltzmann entropy it is by no means
clear that the criticisms levelled at this approach by \citeasnoun{rid&red}
are all valid. In Sec.\ \ref{tge} we described two methods for following
the evolution of the coarse-grained Gibbs entropy, the first, involving
a coarse-graining of the fine-grained distribution at each instant of time,
and the second a sequence of re-coarse-graining as time progresses. The former,
which is the standard understanding of the procedure \cite[p.\ 55]{den&den},
does not yield a strict monotonic increase of entropy. However, it does allow
the system to retrace its steps, either by
velocity reversal or reflection(see  Fig.\ \ref{spin-echofig5}).
This is in conflict with the remarks of \citeasnoun[p.\ 1250]{rid&red}
that a ``reversal of the dynamical evolution in the coarse-grained case does not cause
the distribution to evolve back to its original form''. They appear to be thinking
of the (obvious) impossibility of un-coarse-graining a coarse-grained distribution.
The occurrence of the echo in these circumstances would certainly be
``completely miraculous'' (ibid, p.\ 1251), but this is not how coarse-grained
evolution should be implemented. In any event, the more important question,
raised by \citeasnoun[p.\ 1251]{rid&red}, is whether the spin--echo
system is a ``counterexample to the second law of thermodynamics''.
The answer to this is surely that it depends on what you mean by the second
law of thermodynamics. If, along with Maxwell and Boltzmann and probably
the majority of physicists \citeaffixed[p.\ 113]{rue2}{see e.g.} entropy increase
in an isolated system is taken to be highly probable but not certain, then the spin--echo
model, along with simulations of other simple models \cite{lavis3a}, is a nice
example of the workings of the law. However, if entropy increase is an
iron certainty this example is one, and not a special, example of a violation
of the second law. \citeasnoun[p.\ 1251]{rid&red} assert that the spin--echo
experiments are not a violation of the second law because ``we do not have a
situation where a system evolves spontaneously from a high entropy state to
a low entropy state.'' Apart from the obvious conflict with the quote from
\citename{skl} given above, this, of course, depends on what you mean by ``spontaneous''.
Any experiment or simulation involves preparing the system in some initial state from
which it evolves {\em spontaneously}. There is no conceptual reason why the system cannot be
prepared in a state from which the entropy spontaneously decreases. It just
difficult to do because of their relative paucity. As we have already indicated
in our discussion the best way to find such a state is to let the system
find it itself by evolving in the reverse direction. Then restarting the system
in this state it will show a `spontaneous' decrease in entropy.
\section{Conclusions}
We have considered the case made for the spin--echo experiments being an example
of a special system which destroys the argument for using coarse-graining.
We have argued that the reason for the Boltzmann version of coarse-graining
has nothing to do with the inability to do fine-grained dynamic
calculations,\footnote{Indeed to calculate the Boltzmann entropy one needs
the dynamic detail.} or experiments, but is based on the necessity to
have a demarkation between the micro- and macro-scales. The same arguments
apply to Gibbs-Ehrenfest coarse-graining. The spin--echo experiments are
of technical significance, particularly in respect of the fact that
the echoing procedure can be effected by macroscopic means, but as
a theoretical model of an assembly of non-interacting microsystems
it is in no way special, as we have shown elsewhere \cite{lavis3a}.

\end{document}